\documentclass[manuscript]{emulateapj}

\def\ltsima{$\; \buildrel < \over \sim \;$}
\def\simlt{\lower.5ex\hbox{\ltsima}}
\def\gtsima{$\; \buildrel > \over \sim \;$}
\def\simgt{\lower.5ex\hbox{\gtsima}}
\def\kpc{{\rm\,kpc}}

\def\deg{^\circ}
\def\s{\ifmmode \widetilde \else \~\fi}
\def\={\overline}

\def\spose#1{\hbox to 0pt{#1\hss}}

\def\lta{\mathrel{\spose{\lower 3pt\hbox{$\mathchar"218$}}
     \raise 2.0pt\hbox{$\mathchar"13C$}}}
\def\gta{\mathrel{\spose{\lower 3pt\hbox{$\mathchar"218$}}
     \raise 2.0pt\hbox{$\mathchar"13E$}}}
\def\Dt{\spose{\raise 1.5ex\hbox{\hskip3pt$\mathchar"201$}}}    % upper case
\def\dt{\spose{\raise 1.0ex\hbox{\hskip2pt$\mathchar"201$}}}    % lower case

\def\dotsfill{\leaders\hbox to 1em{\hss.\hss}\hfill}

\def\FeH{{\rm[Fe/H]}}

\shorttitle{Halo stellar structures in the TriAnd region}
\shortauthors{N. Martin, R. Ibata \& M. Irwin}

\begin{document}


\title{Galactic halo stellar structures in the Triangulum-Andromeda region\footnote{Based on observations obtained with MegaPrime/MegaCam, a joint project of CFHT and CEA/DAPNIA, at the Canada-France-Hawaii Telescope (CFHT) which is operated by the National Research Council (NRC) of Canada, the Institut National des Science de l'Univers of the Centre National de la Recherche Scientifique (CNRS) of France, and the University of Hawaii.}}


\author{Nicolas F. Martin}
\affil{Max-Planck Institut f\"ur Astronomie, K\"onigstuhl 17, D-69117 Heidelberg, Germany}
\email{martin@mpia-hd.mpg.de}

\author{Rodrigo A. Ibata}
\affil{Observatoire de Strasbourg, 11 rue de l'Universit\'e, F-67000 Strasbourg, France}
\email{ibata@astro.u-strasbg.fr}

\and

\author{Mike Irwin}
\affil{Institute of Astronomy, Madingley Road, Cambridge CB3 0HA, U.-K.}
\email{mike@ast.cam.ac.uk}

%\altaffiltext{1}{Visiting Astronomer, Cerro Tololo Inter-American Observatory.
%CTIO is operated by AURA, Inc.\ under contract to the National Science
%\altaffiltext{2}{Society of Fellows, Harvard University.}
%\altaffiltext{3}{present address: Center for Astrophysics,
%    60 Garden Street, Cambridge, MA 02138}
%\altaffiltext{4}{Visiting Programmer, Space Telescope Science Institute}
%\altaffiltext{5}{Patron, Alonso's Bar and Grill}

\begin{abstract}
This letter reports on the Galactic stellar structures that appear in the foreground of our Canada-France-Hawaii-Telecopse/MegaCam survey of the halo of the Andromeda galaxy. We recover the main sequence and main sequence turn-off of the Triangulum-Andromeda structure recently found by Majewski and collaborators at a heliocentric distance of $\sim20\kpc$. The survey also reveals another less populated main sequence at fainter magnitudes that could correspond to a more distant stellar structure at $\sim28\kpc$. Both structures are smoothly distributed over the $\sim76\textrm{deg}^2$ covered by the survey although the closer one shows an increase in density by a factor of $\sim2$ towards the North-West. The discovery of a stellar structure behind the Triangulum-Andromeda structure that itself appears behind the low-latitude stream that surrounds the Galactic disk gives further evidence that the inner halo of the Milky Way is of a spatially clumpy nature.
\end{abstract}

\keywords{Galaxy: evolution --- Galaxy: halo --- Galaxy: structure --- Local Group}

\section{Introduction}
The stellar halos of galaxies such as our own Milky Way are expected to have formed over time by the accretion of stellar material from satellites falling within its gravitational potential (e.g. \citealt{freeman02} and references therein). The discovery of the disrupting Sagittarius dwarf galaxy \citep{ibata94} and the stream of stars it leaves in the Galactic halo \citep{ibata01b, ibata02b, majewski03} has revealed that such accretions indeed take place at the present time. Currently, the best evidence for the clumpy nature of the inner halo of the Milky Way ($D\lta50\kpc$) is given by the Sloan Digital Sky Survey (SDSS) whose mapping of the North Galactic Cap shows numerous coherent streams or more fuzzy and diffuse stellar structures believed to be remnants of past accretions \citep{belokurov06b,bell07}.

Aside from large sky surveys such as the SDSS, the search for stellar structures surrounding the Milky Way has also strongly benefited from wide-field mappings of the Andromeda galaxy (M31) and its neighborhood. These mainly photometric surveys need to be deep enough to track M31 red giant branch (RGB) stars and therefore also contain the brighter and bluer main sequence stars that belong to foreground stellar structures. \citet{ibata03} have used their Isaac Newton Telescope (INT) survey to show that the low latitude stream (LLS), first discovered by \citet{newberg02} in the anticenter direction and also referred to as the Monoceros Ring, actually surrounds the Milky Way disk and is present along the M31 line of sight. Kinematics of these main sequence stars also agree with previous detections of this stream, on a roughly circular orbit around our Galaxy \citep{martin06a}. Using ten deeper fields observed within $\sim10\deg$ of M31, \citet{majewski04b} have revealed that another stellar structure can be seen behind the LLS as a fainter and sparser main sequence. A companion paper by \citet{rocha-pinto04} tracks this so-called Triangulum-Andromeda (TriAnd) structure with RGB stars selected in the 2MASS catalogue. Spectroscopic follow-up of these stars unveiled a metallicity $\FeH\sim-1.2$ and a low velocity dispersion consistent with an accretion origin.

In this paper, we take advantage of our deep and wide survey of the Southern region of the M31 outer halo \citep{martin06b,ibata07} to study these foreground stellar structures. The survey was observed with the MegaCam wide-field camera mounted on the Canada-France-Hawaii Telescope and currently represents the deepest wide dataset toward M31. We use it to characterize the TriAnd structure and reveal yet another main sequence fainter than TriAnd and that could correspond to a farther stellar structure. \S~2 briefly presents the dataset while \S~3 studies the two main sequences that appear in the color-magnitude diagram of the MegaCam data. \S~4 concludes this letter.

\section{Data}
The extent of the MegaCam survey studied in the letter is delimited in Figure~1 (thick line polygon). It corresponds to 84\,MegaCam fields observed between 2003-2006 and covers $\sim76$\,sq. deg. Compared to \citet{martin06b} and \citet{ibata07}, six fields that lie close to the M33 galaxy and two fields that overlap with the M31 disk are not considered here. They contain young stars that contaminate the blue regions of the color-magnitude diagram (CMD) where foreground Galactic main sequences (MS) are found. The location of the fields that were used by \citet{majewski04b} to discover TriAnd are also shown Figure~1 as crossed squares.

We refer the reader to \citet[][sections~2 and~3]{ibata07} for a detailed description of the dataset. We would like to emphasize that even at the faint end of the CMD regions studied here ($i\sim23.0$), photometric uncertainties are $\lta0.05$ and completeness is $\gta90$\%; none of these effect should hamper our conclusions. Moreover, given the clear difference between maps of background galaxies and metal-poor M31 stars in the survey (respectively Figures~12 and~20d), star-galaxy separation does not appear to be an issue either.

\begin{figure}
\begin{center}
\includegraphics[angle=270,width=\hsize]{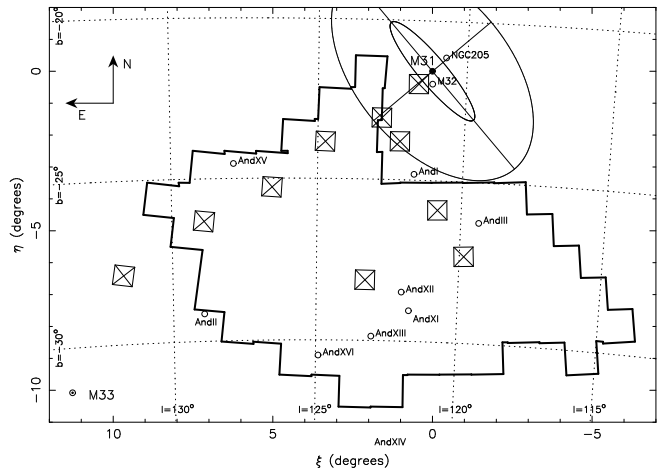}
\caption{Extent of the MegaCam survey in the Southern quadrant of the M31 halo (thick line). North is to the top and East to the left. The dotted lines correspond to a grid in Galactic coordinates. Known M31 satellites are represented by circles or a dotted circle in the case of M33. The inner ellipse around M31 approximately corresponds to its H\textsc{i} disk with a radius of $27\kpc$ while the outer ellipse marks a distance of $50\kpc$ and ellipticity 0.6. The INT survey in which the TriAnd structure is observed \citep{ibata03} fills the outer ellipse. The ten fields that \citet{majewski04b} used to discover the structure are shown as crossed squares.}
\end{center}
\end{figure}

\section{Results}

\begin{figure}
\begin{center}
\includegraphics[width=\hsize]{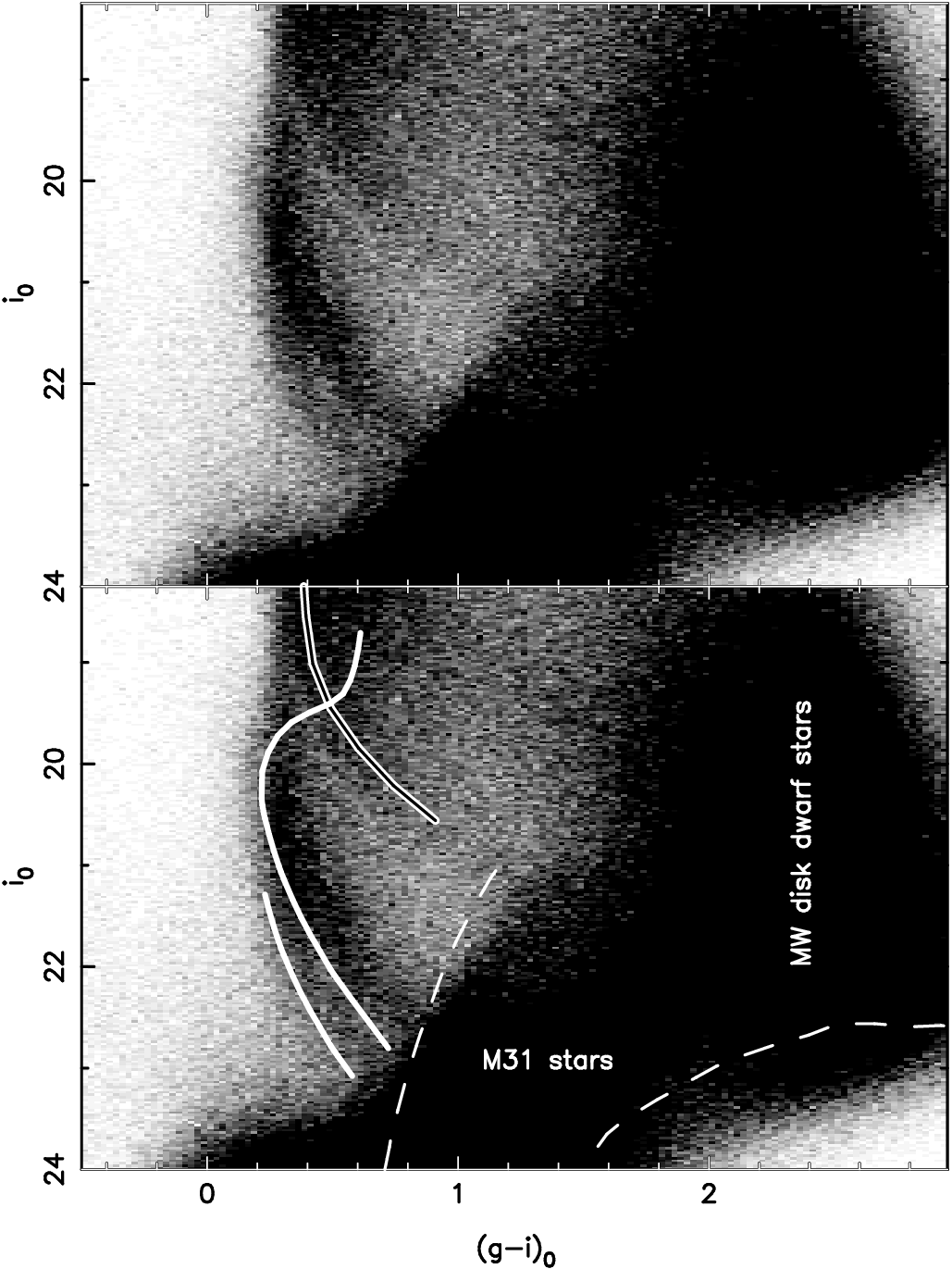}
\caption{Hess diagram of the $\sim1.9\times10^6$ stars in the MegaCam survey. Both panels present the same CMD, with labels and isochrones  overlaid on the bottom panel. The densest features correspond to background M31 stars (between the two dashed isochrones of metallities $\FeH=-2.3$ and $0.0$ placed at the distance of M31) and foreground Milky Way disk stars that pile up at $(g-i)_0\gta1.8$. The main sequence of the TriAnd structure is clearly visible between $(g-i,i)_0\sim(0.3,19.5)$ and $(g-i,i)_0\sim(0.8,22.5)$. Another fainter and more diffuse main sequence is also visible for $i_0\gta21.5$ before it merges with the TriAnd MS. On the bottom panel, the double line corresponds to the LLS main sequence as detected by \citet{ibata03} and the white full lines represent the \citet{girardi04} isochrone of 10\,Gyr and $\FeH\sim-1.3$ shifted to a distance modulus of 16.5 and 17.25 to fit the TriAnd and TriAnd2 MSs, respectively.}
\end{center}
\end{figure}

The Hess diagram of the $\sim1.9\times10^6$ stars found in the MegaCam survey is shown Figure~2. The most obvious feature, located within the two dashed isochrones in the bottom panel is produced by RGB stars at the distance of the Andromeda galaxy. The broadness of this feature is due to the numerous structures with different metallicities that are traced in the halo of M31. These include metal-poor dwarf galaxies such as And~II and~III that create the bluer RGBs and the \citet{ibata01c} metal-rich giant stream that is currently disrupting in the M31 halo and produces the redder RGBs (see \citealt{ibata07} for more details on these structures related to the Andromeda galaxy). Foreground Galactic disk dwarf stars at various distances along the line of sight create the red feature at $(g-i)_0\sim1.8$ and populate the CMD in the $i_0\lta20.0$ region as these stars evolve along their main sequence and become bluer. Finally, the TriAnd MS turn-off clearly appears in the bluer regions of the CMD and extends from $(g-i,i)_0\sim(0.3,19.5)$ down to $(g-i,i)_0\sim(0.8,22.5)$. This main sequence ``forks'' at $i_0\sim21.5$ to reveal another more diffuse main sequence at fainter magnitudes. This fainter MS has to our knowledge never been detected before probably due to its faintness (for instance, the INT survey of the inner halo of M31 in which LLS and TriAnd are detected is not deep enough to show it) and its low density that prevents detections from deep ``pencil-beam'' observations toward M31.

\begin{figure*}
\begin{center}
\includegraphics[angle=270,width=0.24\hsize]{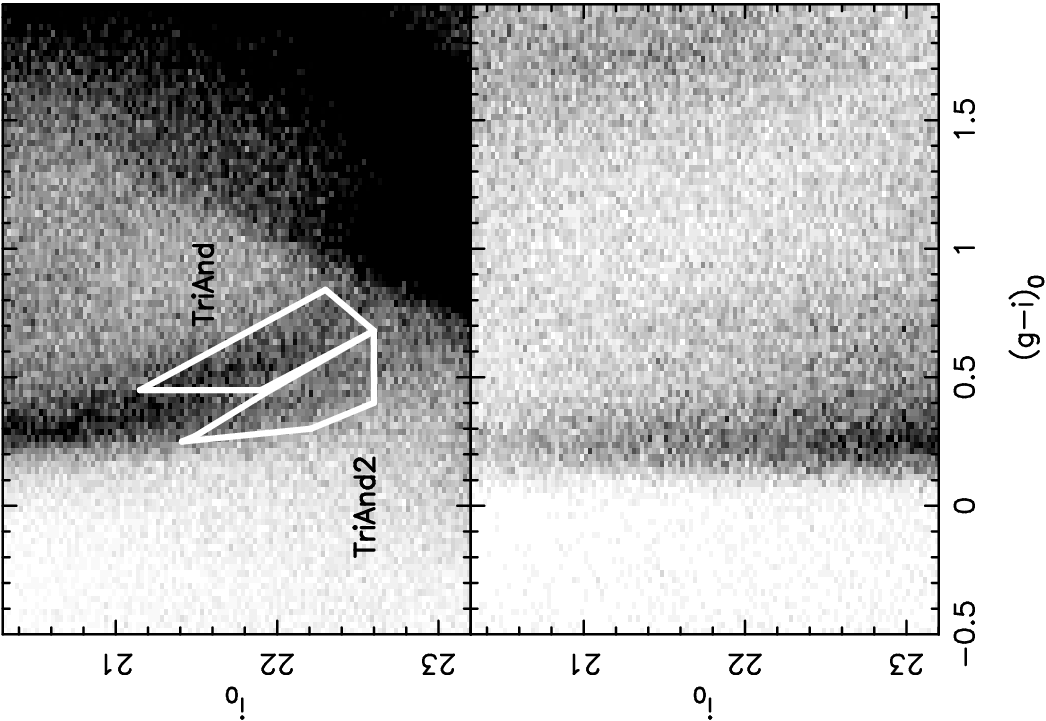}
\includegraphics[angle=270,width=0.37\hsize]{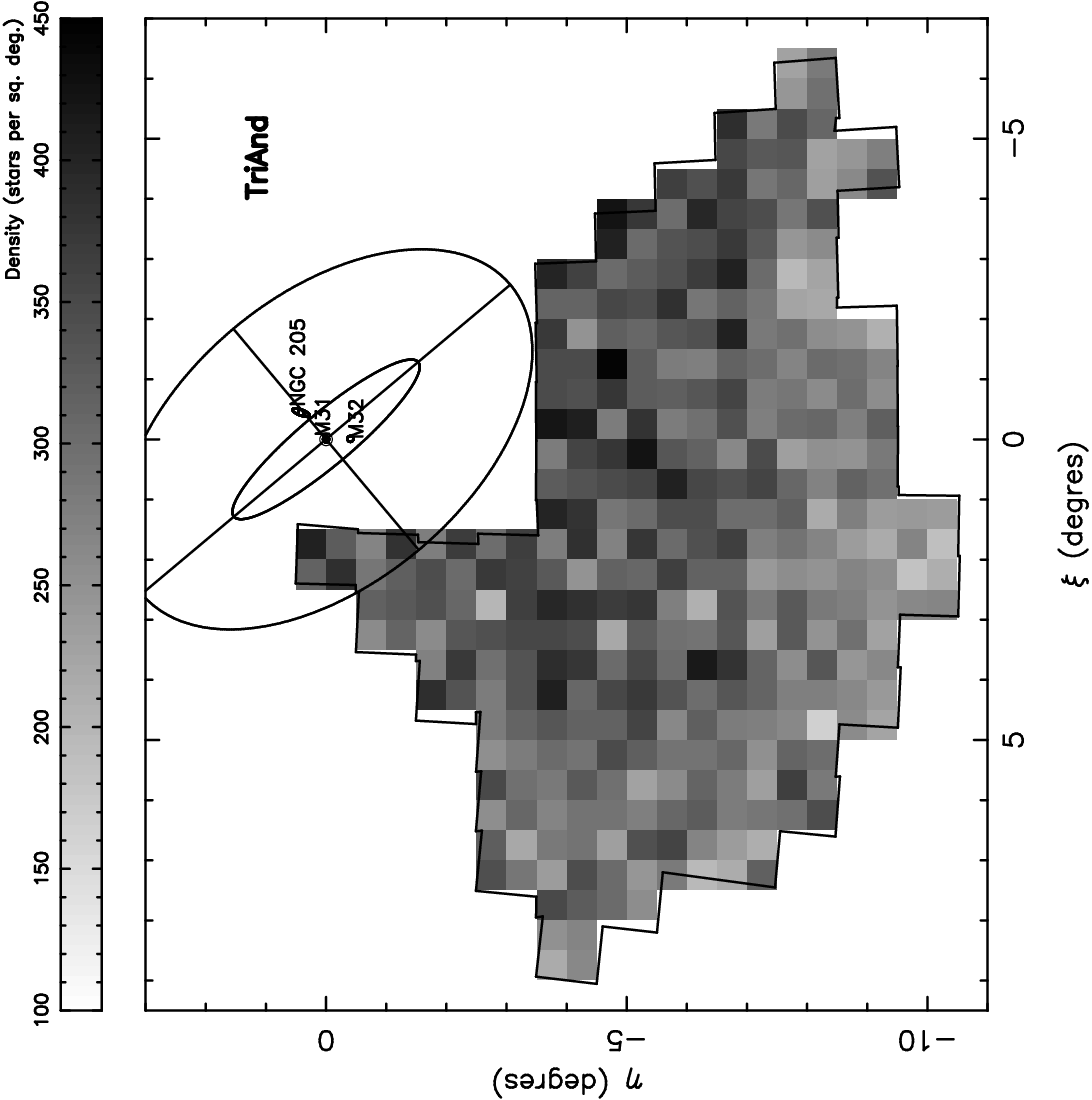}
\includegraphics[angle=270,width=0.37\hsize]{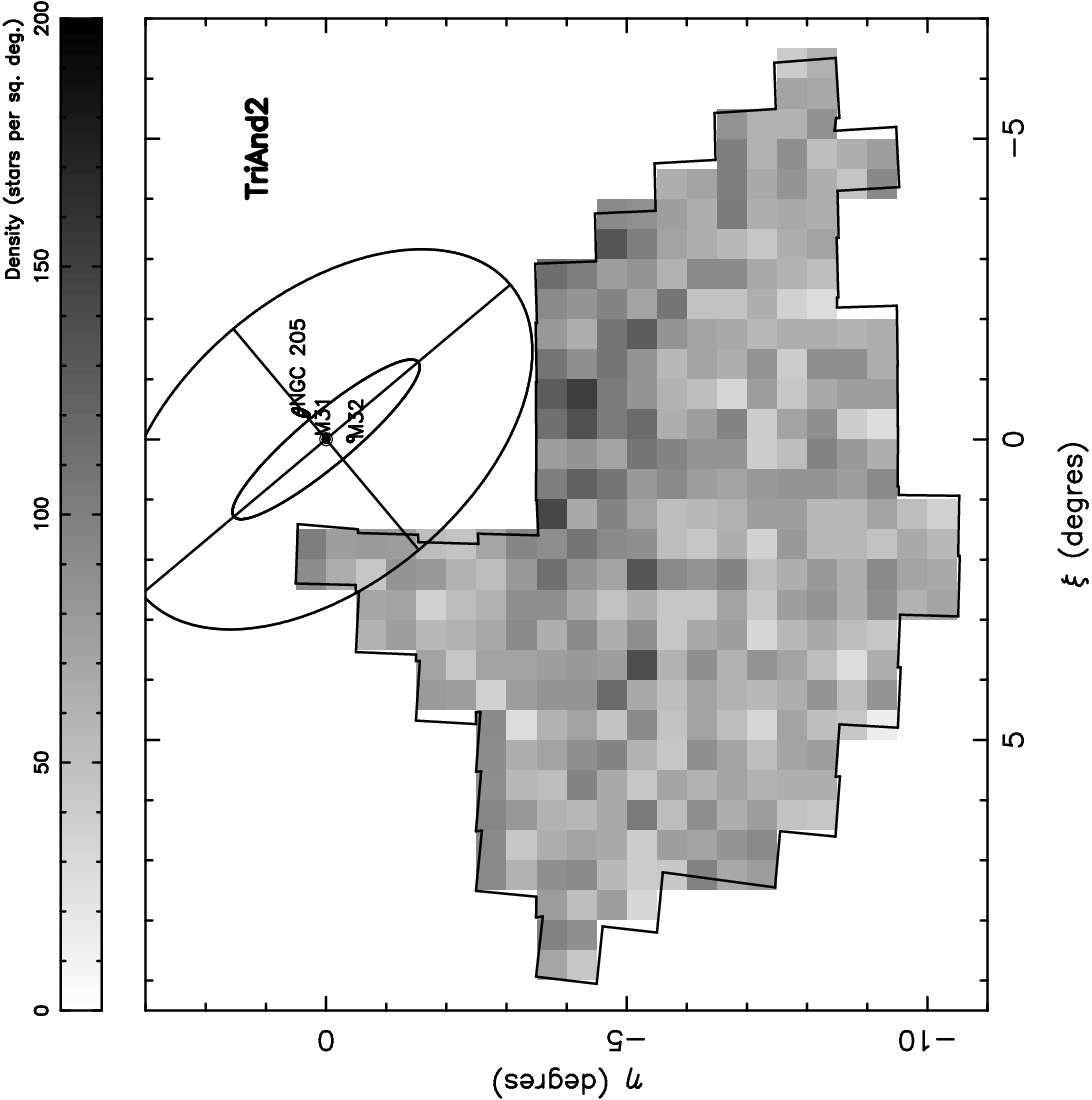}
\caption{CMDs of the MegaCam survey and Besan\c{c}on model (top- and bottom-left respectively) with the selection boxes used to construct the density maps of the TriAnd (middle) and TriAnd2 (right) structures. Pixels that are more than 50 percent within the boundaries of the MegaCam survey are corrected from incompleteness. Pixels with a lower completeness factor are not shown. TriAnd shows denser regions in the North-Western parts of the survey ($\xi\simlt4\deg$ and $\eta\simgt-7\deg$) whereas TriAnd2 appears smoother. None of the two structures show a well-defined core or center.}
\end{center}
\end{figure*}

The ridge of the LLS main sequence discovered by \citet{ibata03} in their INT survey has been plotted in the bottom panel of Figure~2. It overlaps reasonably well a denser feature of the CMD produced by stars in the Northern-most fields of the survey, where the LLS was tracked in the INT data (down to $b\sim-24\deg$). The TriAnd MS is $\sim2$\,magnitudes fainter than the LLS fiducial but the bluer turn-off of TriAnd ($(g-i)_0\sim0.15$) compared to LLS ($(g-i)_0\sim0.4$) probably indicates a difference in age and/or metallicity between the two structures and therefore precludes a simple shift of the LLS fiducial to determine the distance to TriAnd. However, using the spectroscopic metallicity measured for TriAnd RGB stars ($\FeH\sim-1.2$; \citealt{rocha-pinto04}), one can constrain the age and distance of the structure from the color and magnitude of the turn-off. It can be seen on the right panel of Figure~2 that the \citet{girardi04} isochrone with a metallicity $Z=0.001$ (corresponding to $\FeH\sim-1.3$), an age of 10\,Gyr and a distance modulus of 16.5 gives a good agreement with the MegaCam observations. This corresponds to a heliocentric distance of $\sim20\kpc$ and a Galactocentric distance of $\sim25\kpc$. Although the turn-off of the fainter MS is hidden behind the TriAnd MS and MS turn-off, the same isochrone provides a good fit when assuming a distance modulus of 17.25. This converts to heliocentric and Galactocentric distances of $\sim28\kpc$ and $\sim33\kpc$ respectively. For clarity, we will call this new feature TriAnd2 hereafter.

Since there is no region in the MegaCam survey where these CMD features seem absent, we rely on the Besan\c{c}on model \citet{robin03} to check whether they are expected from Milky Way components (disk and mainly stellar halo in these blue regions). We extract a 25 sq. deg. region from the model with $-30\deg<b<-25\deg$ and $125\deg<l<130\deg$ and apply the same smoothing as in \citet{ibata07} to account for the unphysical sharpness of the CMD structures in the model (see their Figure~15). Completeness is not corrected in this letter since it does not have a significant impact in the region of the CMD we are interested in. The survey and the model CMDs are shown in the top- and bottom-left panels of Figure~3, normalized to 1 sq. deg. to allow a proper comparison. They have dramatically different features: the model does not show any of the two MSs but its stellar halo produces a blue vertical feature at $(g-i)_0\sim0.25$ that is not present (or at least not as dense) in the survey. Although TriAnd is a clear $2-3\times$ overdensity compared the model, this is not the case of TriAnd2 that is much fainter and falls in the CMD region where the density of halo stars increases in the model. But no turn-off is observed in the model CMD where halo stars never bend to redder colors as the TriAnd2 MS does (Figure~2). Interestingly, a comparison of the COSMOS data ($l=237\deg$, $b=+42\deg$) with the Besan\c{c}on model reveals that a ``hook-like'' feature very similar to the MSs observed toward M31 may be due to a stellar structure that inhabits the Milky Way halo at $22-34\kpc$ in this direction \citep{robin07}. Therefore, it appears more likely that the TriAnd2 CMD is indeed a MS, that it is not linked to a smooth Galactic components (disk or stellar halo) and belongs to a stellar structure that inhabits the inner halo of the Milky Way. This adds to the growing evidence that the Milky Way halo is not well modeled by a simple smooth model \citep{bell07}.

Density maps of TriAnd and TriAnd2 stars are also shown Figure~3, with stars selected along the two main sequences, within the selection boxes overlaid on the top-left CMD of the Figure. These two boxes are chosen independently to maximize the number of selected MS members while reducing contamination by avoiding M31 stars and the region where the 2 features merge in the CMD. As such, the map densities are not directly comparable but provide a tool to study the spatial evolution of both stellar structures over the area of the survey. The map of metal-poor M31 stars (Figure~20d of \citealt{ibata07}) is very distinct from the TriAnd/TriAnd2 maps and shows that there is no significant M31 contamination of the maps of Figure~3. If such a contamination was present, the numerous metal-poor streams found by in the East of the survey would also clearly appear here. TriAnd shows some fluctuations in its density, with an approximately twofold increase\footnote{This becomes a threefold increase when assuming the $\sim100$ stars per sq. deg. from the Besan\c{c}on model in the same selection box is a reasonable estimate of the underlying CMD contamination.} between the low density Southern region and the higher density North-Western region of the survey ($\eta\simgt-7\deg$ and $\xi\simlt4\deg$). TriAnd2 on the other hand shows no significant variation in its density over the survey.

\begin{figure}
\begin{center}
\includegraphics[angle=270,width=\hsize]{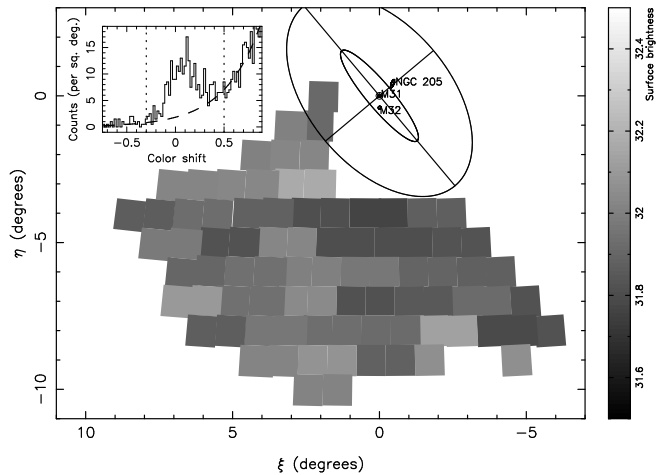}
\caption{Surface brightness map of the combined TriAnd and TriAnd2 features (assuming the TriAnd distance) from stars in the $21.0<i_0<22.0$ magnitude range. The color shift of each star from the TriAnd isochrone (Figure~2) is determined and regions away from the main sequence (delimited by the dotted lines) are fitted by an exponential function (dashed line) to estimate the underlying background CMD contamination.}
\end{center}
\end{figure}

Although the selection boxes that are used to construct these maps do not allow a simple transformation to surface brightness given the absence of knowledge on the CMD contamination, this can be done in the region of the CMD were the two MSs have merged. We determine a bright limit of the TriAnd surface brightness by assuming that all the stars belong to the closer TriAnd MS and, since TriAnd2 does not appear to fluctuate significantly over the survey, variations should reflect those of TriAnd alone. The magnitude range used is $21.0<i_0<22.0$ and corresponds to $4.7<M_V<5.8$. According to the \citet{dotter07} luminosity function for a population of 10\,Gyr, $\FeH=-1.5$ and a Salpeter IMF \citep{salpeter55}, 11.7\% of the luminosity of the stellar population is measured in this magnitude range\footnote{Note that the \citet{silvestri98} luminosity function used by \citet{majewski04b} has a normalization of 16\% in that range and would lead to slightly fainter surface brightness values by $\sim0.15$\,mag}. For each couple of nearby fields (the signal is too low in single fields), we construct the distribution of stars in this range as a function of their color shift from the TriAnd isochrone of Figure~2. An example of the corresponding histogram is shown in the inset panel of Figure~4 with the stellar overdensities clearly visible. To correct for the underlying CMD contamination we fit an exponential function (the dashed lines) to the regions surrounding the overdensity ($\textrm{color shift} <-0.3$ and $0.5<\textrm{color shift}<0.8$). All counts above the dashed line in the color shift range between $-0.3$ and 0.5 are finally translated to surface brightness in the V band. The map of Figure~4 presents the fluctuations of $\Sigma_V$  in the survey and confirms the findings of the TriAnd counts map, with a $\sim0.3$ magnitude shift between the Southern regions ($\sim32.1$mag./sq. arcsec) and the North-Western ones ($\sim31.8$mag./sq. arcsec). This corresponds to an increase by a factor of two in the number of stars in the selection box. These values are also in reasonable agreement with the bright limits measured in a similar way by Majewski et al. (2004; $\Sigma_V=32.0$ or 32.4 depending on their isochrone assumptions). However, it does not take into account the merging of the two MSs in the region of the CMD that was used. For instance, if 25\% of the stars in fact belong to the fainter TriAnd2 main sequence, the surface brightness of TriAnd would be fainter by $\sim0.3\,\textrm{mag}/\textrm{arcsec}^2$.

%Since TriAnd2 is very elusive and it does not show evidence of density fluctuation over the $\sim18\kpc^2$ covered by the MegaCam survey at this distance, it is unclear whether is corresponds to the stellar halo of the Milky Way or whether it is a genuine stellar structure that is observed in addition to the halo. As no other survey can be used for comparison, we use the Besan\c{c}on model to test the expected behavior of the halo (although the stellar halo is not well constrained in the model \citealt{robin00}). Simple star counts in the selection box used to construct the TriAnd2 map give $\sim150/sq.deg.$ at the region $120\deg<l<125\deg$ and $-30\deg<b<-25\deg$, a value that is slightly higher than that found in the survey ($\sim100/sq.deg.$) and would lead to think there is no need to invoke another stellar structure around the Milky Way. However, no turn-off is observed in the color-magnitude diagram of the model as stars belonging to the stellar halo form a straight blue edge at $(g-i)_0\sim0.2$ and never bend to redder colors as the MSs do in Figure~2. Interestingly, a comparison of the COSMOS data with the Besan\c{c}on model reveals that a ``hook-like'' feature very similar to the MSs observed towards M31 may be due to a stellar structure that inhabits the Milky Way halo in this direction \citep{robin07}. Therefore, it appears more likely that the TriAnd2 CMD is indeed a MS, that it is not linked to a smooth Galactic components (disk or smooth stellar halo) and belongs to a stellar structure that inhabits the inner halo of the Milky Way. 

\section{Discussion}
The presence of two Galactic stellar structures in the MegaCam data at Galactocentric distances of $\sim25$ and $\sim33\kpc$, added to the low-latitude stream observed only a few degrees away \citep{ibata03} shows the clumpy nature of the inner halo of the Milky Way in this region. But since these structures do not show any boundary within the $\sim76\,\textrm{deg}^2$ covered here, there is no reason that this conclusion should only be applied toward M31. In fact, each step forward in the depth and/or coverage of the inner halo reveals hitherto undetected stellar structures within a few tens of kiloparsecs (e.g. \citealt{belokurov06b}, \citealt{robin07}). This increasingly clumpy and complex nature of the inner halo of the Galaxy is reminiscent of the inner halo of M31 which also contains large amounts of substructure with much of it in the form of a large extended rotating component \citep{ibata05}. 

However, this does not rule out that some of these stellar structures could be linked together. It has already been proposed by \citet{majewski04b} that the TriAnd structure could be another loop of the low latitude stream, produced by its progenitor spiraling toward the Galactic center. Simulations of the stream by \citet{penarrubia05} or \citet{martin05b} indeed show that TriAnd is a natural outcome of such an accretion. Could the new structure found here also be related to TriAnd? The good fit of the same isochrone to both main sequences in Figure~2 indicates a similar stellar population. However, the turn-off of this structure is bluer than that of the LLS and would argue against a link, although it could also indicate that the putative progenitor of all these structures contained a stellar population gradient. A definite answer will have to wait for spectroscopic observations of the new structure to determine from their radial velocities if they are on orbits compatible with TriAnd and the low-latitude stream.

\acknowledgments

We are very grateful to the CFHT staff for performing the MegaCam observations in queue mode. NFM thanks Eric Bell for comments on this letter.

{\it Facilities:} \facility{CFHT (MegaCam)}.

\bibliographystyle{apj}

% Bibtex will create a .bbs file in the directory and before sending to the editor, I should replace the bibliography call by this file.

\clearpage

\clearpage

\end{document}